\documentclass[a4paper,12pt]{article}
\usepackage{amsmath,amssymb,bm}
\usepackage{mathrsfs}
\usepackage[dvipdfmx]{color,graphicx}
\usepackage{enumerate}

\title{On the Approximate Purification of Mixed Strategies in Games with Infinite Action Sets}
\author{Yuhki Hosoya\thanks{Corresponding Author, E-mail:hosoya(at)tamacc.chuo-u.ac.jp, ORCID ID:0000-0002-8581-4518}~{}\thanks{We are grateful to Toru Maruyama, Shinsuke Nakamura, Takako Fujiwara-Greve, and Toru Hokari for their helpful comments and suggestions. We also thank the anonymous reviewers for their helpful comments and suggestions on this paper. This work was supported by JSPS KAKENHI Grant Number JP21K01403.}\\
Faculty of Economics, Chuo University\\
Chaowen Yu\\
Department of Economics, Keio University}
\date{}
\newcommand{\ve}{\varepsilon}
\begin{document}
\maketitle

\begin{abstract}
We consider a game in which the action set of each player is uncountable, and show that, from weak assumptions on the common prior, any mixed strategy has an approximately equivalent pure strategy. The assumption of this result can be further weakened if we consider the purification of a Nash equilibrium. Combined with the existence theorem for a Nash equilibrium, we derive an existence theorem for a pure strategy approximated Nash equilibrium under sufficiently weak assumptions. All of the pure strategies we derive in this paper can take a finite number of possible actions.

\vspace{12pt}
\noindent
{\bf JEL codes}: C62, C65, C72, C73.

\vspace{12pt}
\noindent
{\bf Keywords}: mixed strategy, approximate purification, uncountable action set, conditionally atomless, Nash equilibrium.
\end{abstract}

\section{Introduction}
The notion of the mixed strategy was introduced by von Neumann and Morgenstern (1944) to prove the minimax theorem, and was later used in Nash's (1950) existence theorem for a Nash equilibrium. In game theory, there are many cases in which the existence of a solution can be guaranteed using the concept of mixed strategies. Thus, this concept continues to be extensively used today.

Mixed strategies have, however, been criticized for a long time because, if a player wants to implement a mixed strategy, he/she has to prepare his/her own randomizing device. From a positive point of view, a solution is interpreted to represent human behaviors in the real world. However, there is almost no behavior in the real world that can be interpreted as that of using a randomizing device. From a normative point of view, a solution to a game is considered to be a proposal presented by the mediator of the conflict. However, a proposal that would require each party to use a randomizing device to resolve a dispute would be difficult for a mediator to submit and, even if submitted, is unlikely to be accepted by each party. In conclusion, solutions derived using mixed strategies are, in many cases, problematic.

On the other hand, in the context of Bayesian games, a strategy is described as a function from the space of outside signals into the space of actions. This strategy is randomized by the signal, and can take probabilistic actions without the need to use a randomization device. As a result, the problem described above is not encountered. Thus, if we can create a ``pure strategy'' that is ``equivalent'' to a given mixed strategy, the problem of interpretation will disappear. The research theme of {\bf purifying mixed strategies} was born from this idea.

In this research area, it is known that a certain independence in the signal structure results in the existence of a pure strategy that is ``exactly equivalent'' to a given mixed strategy (e.g., Radner and Rosenthal (1982), Milgrom and Weber (1985)). These results have been treated in a unified manner by Khan et al. (2006). In these studies, the set of possible actions were assumed to be finite. Khan and Rath (2009) extended this result to the case where the action set is countably infinite. However, this result cannot be extended to the case where the set of possible actions is uncountable, as in the case of Cournot games. In this case, Khan and Zhang (2014) produced the following interesting result: in games where the set of actions for each player has a continuum cardinality, a necessary and sufficient condition for there to have to be a pure strategy Nash equilibrium that is ``exactly equivalent'' to a given mixed strategy Nash equilibrium is that the space of signals is {\bf saturated}.

We view this result of Khan and Zhang (2014) as a kind of impossibility theorem. To guarantee the existence of a pure strategy that is equivalent to a mixed strategy, the space of signals must be saturated. As we argue in Section 4 of this paper, the assumption of saturation implies that the measurable structure of the space of signals is ``very fine'' in some sense, meaning that an incredible variety of randomness can be extracted from the signals. However, in a phenomenon that can be viewed as a game-theoretic situation, it is unlikely that players are able to receive such many outside signals. Additionally, in many applied studies, a Polish space with a Borel probability is adopted as the space of signals. However, {\bf any Polish space with a Borel probability is not saturated}. Therefore, it is impossible to show the existence of an ``exactly equivalent'' pure strategy under the usual assumptions.

In this study, we consider another approach: the existence of a pure strategy that is ``approximately equivalent'' to a given mixed strategy. Aumann et al. (1983) showed that there exists a pure strategy that is ``approximately equivalent'' to an arbitrary given mixed strategy when only certain non-atomic properties of the common prior are assumed. This result has the weakness that it can only guarantee the existence of an ``approximately equivalent'' pure strategy, but has the advantage that it does not require independence in the signal structure and guarantees the existence of such a strategy in very general situations. Note that the non-atomic property is also needed for ``exact equivalence'' results, and thus the assumption required for this result is weaker than that required for ``exact equivalence'' results. Although the action set is assumed to be finite in Aumann et al. (1983), we extend this result to the case where the action set is an arbitrary compact metric space (Theorem 1, Theorem 2). Our assumption allows that the space of signals is a Polish space with a Borel probability, and thus it is not necessarily assumed to be saturated. In this sense, we consider that this result is a non-trivial generalization of past studies.

The assumption regarding the common prior can be further weakened when considering the purification of a Nash equilibrium. This result was also shown by Aumann et al. (1983) for the case where each action set is finite, but this study extends this result to the case where each action set is a compact metric space (Theorem 3). Balder (1988) proved the existence theorem for a Nash equilibrium with mixed strategies under certain assumptions. Connecting this result with ours, we can show the general existence of an approximated Nash equilibrium with pure strategies (Theorem 4). This result is an extension of Corollary of Milgrom and Weber (1985).

Note that all pure strategies derived in our results can be considered as functions that take values in a finite subset of the action set. This is one of the good properties of our results, as real people are not always able to finely change their behavior in response to signals.

In subsection 2.1, we introduce our model and several seminal results. In subsection 2.2, we introduce the requirements for the common prior used in this study. In subsection 2.3, we explain our setup of the space of mixed strategies. Section 3 then presents our main results. In Section 4, we mention several related works, and explain how our research can be positioned within the related literature. Section 5 presents the conclusion. The proofs of all results are given in Section 6.

\section{The Model}

\subsection{Basic Notation}
A game considered in this study is represented by a tuple
\[G=(N,(K_i)_{i\in N},(\Omega,\mathscr{S},\mu),u,(X_i,\mathscr{S}_i,x_i)_{i\in N}).\]
The set $N=\{1,...,n\}$ is the set of all players, and $K_i$ is the {\bf action set} for player $i$. We assume that each $K_i$ is a compact metric space. The probability space $(\Omega,\mathscr{S},\mu)$ is interpreted as the space of {\bf outside signals}. The probability measure $\mu$ denotes the {\bf common prior}. The function $u:\prod_{i=1}^nK_i\times \Omega\to \mathbb{R}^m$ is the {\bf payoff function}. Although it is natural that $n=m$ and $u_i$ denotes the payoff of the $i$-th player, we consider that $m$ may be not equal to $n$ for a technical reason. If $m=n$, then we say that this game is {\bf usual}. We assume that $u(k,\omega)$ is continuous in $k$ and measurable in $\omega$, and that there exists a positive integrable function $r:\Omega\to \mathbb{R}$ such that $\|u(k,\omega)\|\le r(\omega)$ for all $(k,\omega)$.

The measurable space $(X_i,\mathscr{S}_i)$ is the {\bf observation space} for player $i$, and the function $x_i:\Omega\to X_i$ is the {\bf observation function}. We assume that, for each $i$, $(X_i,\mathscr{S}_i)$ is isomorphic to $([0,1],\mathscr{F})$, where $\mathscr{F}$ is a $\sigma$-algebra that includes all Borel sets,\footnote{That is, $\mathscr{F}$ is finer than or equal to the Borel $\sigma$-algebra of $[0,1]$.} and $x_i:\Omega\to X_i$ is measurable. Note that, by Kuratowski's theorem,\footnote{See Theorem 3.3.13 of Srivastava (1998).} every uncountable Polish space with the Borel $\sigma$-algebra satisfies the above requirement, and thus our requirement for $(X_i,\mathscr{S}_i)$ is not so strong. Define a function $\mu'$ on $\prod_{i=1}^nX_i$ as follows:
\[\mu'(A)=\int 1_{\{x\in A\}}d\mu.\]
We can easily confirm that $\mu'$ is a probability measure on $\prod_{i=1}^nX_i$. By using $\mu'$ instead of $\mu$, we can assume, without loss of generality, that $\Omega=\prod_{i=1}^nX_i$, $\mathscr{S}=\otimes_{i=1}^n\mathscr{S}_i$, and $\mu$ itself is a probability measure on this $\Omega$. Clearly, in this case $x_i(\omega_1,...,\omega_n)=\omega_i$ for each $i$. In this connection, we omit the notation $\mathscr{S}$ and $(X_i,\mathscr{S}_i,x_i)$, and simply write this game $G$ as $(N,(K_i)_{i\in N},(\Omega,\mu),u)$.

We call a measurable function $f_i:X_i\to K_i$ a {\bf pure strategy} of player $i$. Despite the name, the actual action of player $i$ while following a pure strategy is randomized by the observation of the outside signal $x_i$. Let $\mathscr{M}^1_+(K_i)$ be the set of all probability measures on $K_i$. We call a weakly measurable function\footnote{The definition of weak measurability is explained in subsection 2.3.} $f_i:X_i\to \mathscr{M}^1_+(K_i)$ a {\bf mixed strategy}.\footnote{This name may appear odd. Actually, this function is sometimes called by different names in this research area. For example, Radner and Rosenthal (1982) called such a function a `behavior strategy'. Milgrom and Weber (1985) called it a `behavioral strategy' and discussed the relationship between this and the `distributional strategy' that they used. Balder (1988) called it a `transition probability'.  Aumann et al. (1983) called this function a `mixed strategy'. The name here follows that of Khan and Zhang (2014).} Because $k_i\in K_i$ is identified with the Dirac measure $\delta_{k_i}\in \mathscr{M}^1_+(K_i)$, any pure strategy is also a mixed strategy.\footnote{Later, we will show rigorously that our pure strategy is actually weakly measurable.} $f=(f_1,...,f_n)$ is called a {\bf strategy profile} if each $f_i$ is a (pure or mixed) strategy of player $i$. To avoid confusion, we use the notation $f_{i,x_i}$ instead of $f_i(x_i)$. The expected payoff function $U$ is defined by
\[U(f)=\int ud(f_{1,x_1}\otimes...\otimes f_{n,x_n})d\mu.\]
Using Theorem 2.8 of Billingsley (1999) and applying our Lemma 1 discussed later, we can easily show that $U$ is a well-defined $\mathbb{R}^m$-valued function.

Let $f_i$ be a mixed strategy of player $i$. A strategy $f_i'$ is {\bf exactly equivalent} to $f_i$ if and only if, for every strategy profile $f_{-i}$ of players other than player $i$ and every $j\in \{1,...,m\}$,
\[U_j(f_i,f_{-i})=U_j(f_i',f_{-i}).\]
Similarly, a strategy $f_i'$ is {\bf \mbox{\boldmath $\ve$}-equivalent} to $f_i$ if and only if for every strategy profile $f_{-i}$ of players other than player $i$,
\[\max_{j\in \{1,...,m\}}|U_j(f_i,f_{-i})-U_j(f_i',f_{-i})|<\ve.\]
A pure strategy $f_i'$ is called an {\bf exact purification} (resp. $\ve$-purification) of the strategy $f_i$ if $f_i'$ is exactly equivalent (resp. $\ve$-equivalent) to $f_i$.

\subsection{Conditionally Atomless and Weakly Conditionally Atomless Priors}
Recall that each $X_i$ is a measurable space, $\Omega=\prod_{i=1}^nX_i$, and $\mu$ is a probability measure on $\Omega$. We say that $\mu$ is {\bf conditionally atomless for player \boldmath{$i$}} if the conditional probability $\mu(\cdot|x_{-i})$ is non-atomic almost surely with respect to the marginal probability $\mu_{X_{-i}}$, where $x_{-i}=(x_1,...,x_{i-1},x_{i+1},...,x_n)$ and $X_{-i}=\prod_{j\in N, j\neq i}X_j$ as usual.\footnote{For rigorous definitions of marginal probability and conditional probability, see section 10.2 of Dudley (2002).} We also say that $\mu$ is {\bf weakly conditionally atomless for player \boldmath{$i$}} if for every $j\in N$ such that $i\neq j$, $\mu_{ij}$ is conditionally atomless for player $i$, where $\mu_{ij}(A)=\mu(A\times \prod_{p\in N\setminus \{i,j\}}X_p)$ for every measurable set $A\subset X_i\times X_j$.

It is known that every conditionally atomless probability for player $i$ is weakly conditionally atomless for player $i$, and that the converse is not true. See Proposition 3 and Example 2 of Yu et al. (2017) for detailed arguments. Meanwhile, if $\mu$ is not conditionally atomless, then there may exist a mixed strategy with no $\ve$-purification for sufficiently small $\ve>0$ even when $K_i$ is finite. Therefore, the conditional atomless requirement is crucial for $\ve$-purification.

\subsection{Basic Knowledge in the Space of Probability Measures}
In this section, we present some basic knowledge regarding the space of probability measures on a separable and complete metric space. All of the facts mentioned in this section are proved in ch.11 of Dudley (2002), ch.1 of Billingsley (1999), or ch.1-2 of Parthasarathy (2014).

Let $K$ be a separable and complete metric space. Then, $\mathscr{M}^1_+(K)$ denotes the set of all Borel probability measures on $K$. Recall the definition of the Prohorov metric: for a set $A\subset K$ and $\ve>0$, let $A^{\ve}=\{x\in K|\exists y\in A\mbox{ s.t. }d(x,y)<\ve\}$, wherer $d$ denotes the metric of $K$. For any $P,Q\in \mathscr{M}^1_+(K)$, define
\[\rho(P,Q)=\inf\{\ve>0|P(A)\le Q(A^{\ve})+\ve\mbox{ for each Borel set }A\}.\]
This function $\rho$ is called the {\bf Prohorov metric}. Under this metric, $\mathscr{M}^1_+(K)$ is separable and complete. Moreover, the convergence of $(P^{\nu})$ to $P$ with respect to the Prohorov metric is equivalent to the weak* convergence.\footnote{Recall that $P^{\nu}$ converges to $P$ with respect to the weak* topology if and only if $\lim_{\nu\to \infty}\int udP^{\nu}=\int udP$ for every continuous and bounded function $u:K\to \mathbb{R}$.} Furthermore, there exists a countable dense set $\mathscr{P}\subset \mathscr{M}_1^+(K)$ such that each $P\in \mathscr{P}$ has a finite support, and if $K$ is compact, $\mathscr{M}_1^+(K)$ is also compact with respect to the Prohorov metric.

Let $X$ be some measurable space. Then, we call a function $f:X\to \mathscr{M}^1_+(K)$ {\bf weakly measurable} if, for every continuous function $v:\mathscr{M}^1_+(K)\to \mathbb{R}$, the composition $v\circ f:X\to \mathbb{R}$ is measurable. Note that if $g:X\to K$ is measurable, then a function $f:x\mapsto \delta_{g(x)}$ is weakly measurable, where $\delta_{g(x)}$ is the Dirac measure. Indeed, if we define $h(k)=\delta_k$, then $f(x)=(h\circ g)(x)$. It is easy to show that the function $h$ is continuous, and thus for every continuous function $v:\mathscr{M}^1_+(K)\to \mathbb{R}$, $v\circ f=(v\circ h)\circ g$ is measurable. Thus, we have that $f$ is weakly measurable. In this connection, we have that every pure strategy is also a mixed strategy.

If $f(X)$ is finite and $f^{-1}(P)$ is measurable for every $P\in\mathscr{M}^1_+(K)$, then we call $f$ a {\bf simple function}. 

We can obtain the following result. This is the basis of this paper.

\vspace{12pt}
\noindent
{\bf Lemma 1}. Suppose that $K$ is a separable complete metric space, and $X$ is a measurable space. Let $\mathscr{P}$ be a countable dense set in $\mathscr{M}^1_+(K)$ with respect to the Prohorov metric $\rho$. Then, for every weakly measurable function $f:X\to \mathscr{M}^1_+(K)$, there exists a sequence $(f_{\nu})$ of simple functions such that the range of $f_{\nu}$ is in $\mathscr{P}$ for every $\nu$, and $\rho(f_{\nu}(x),f(x))\to 0$ as $\nu\to \infty$ for every $x\in X$.

\vspace{12pt}
The next lemma asserts the equivalence between our mixed strategy and the `transition probability' used in Balder (1988). Later, we will use Balder's theorem to prove our Theorem 4, and thus this fact is needed. In this lemma, we use the notation $f_x$ instead of $f(x)$ to avoid confusion.

\vspace{12pt}
\noindent
{\bf Lemma 2}. Suppose that $K$ is a compact metric space, and $X$ is a measurable space. Then, a function $f:X\to \mathscr{M}^1_+(K)$ is weakly measurable if and only if for each Borel set $B$ in $K$, the function $x\mapsto f_x(B)$ is measurable on $X$.

\section{Results}
\subsection{First Result: the Existence of an Approximate Purification}
In this section, we assume that $N=\{1,2\}$, and use the following notation: $X_1=X, K_1=K, X_2=Y, K_2=L$. By assumption, $K$ and $L$ are compact metric spaces, and thus are separable and complete. Recall the definition of $\ve$-purification. That is, if $f$ is a mixed strategy of player 1, then a pure strategy $f'$ of player 1 is an $\ve$-purification of $f$ if and only if for every mixed strategy $g$ of player 2,
\[\max_{i\in \{1,...,m\}}|U_i(f,g)-U_i(f',g)|<\ve.\]
Our first main result is as follows.

\vspace{12pt}
\noindent
{\bf Theorem 1}. Suppose that $N=\{1,2\}$ and $\mu$ is conditionally atomless for player 1. Then, for every $\ve>0$ and every mixed strategy $f$ of player 1, there exists a finite subset $K'$ of $K$ and a $\ve$-purification $f'$ of $f$ such that for every $x$, $f_x'$ is included in $K'$.

\vspace{12pt}
As a corollary, we obtain the following result.

\vspace{12pt}
\noindent
{\bf Theorem 2}. Suppose that $N=\{1,...,n\}$ and $\mu$ is conditionally atomless for player $i$. Then, for every $\ve>0$ and every mixed strategy $f_i$ of player $i$, there exists a finite subset $K_i'$ of $K_i$ and an $\ve$-purification of $f_i'$ such that for every $x_i$, $f_{i,x_i}$ is included in $K_i'$.

\subsection{Second Result: the Existence of a Pure Approximated Nash Equilibrium}
We should define the notion of approximated Nash equilibria. Recall that a game is said to be usual if and only if the dimension of the range of $U$ is the same as the number of players. Suppose that the game is usual. Then, the function $U_i$ can be seen as the payoff of player $i$. For a given $\ve\ge 0$, the strategy profile $f$ is an $\ve$-Nash equilibrium if and only if, for every strategy $g_i$ of player $i$,
\[U_i(g_i,f_{-i})\le U_i(f)+\ve.\]
We call a $0$-Nash equilibrium a Nash equilibrium as usual. If $f$ is a Nash equilibrium, then $f'$ is called an $\ve$-purification of $f$ if the following requirements holds.
\begin{enumerate}[1)]
\item For every $i$, $f_i'$ is a pure strategy of player $i$.

\item If $f''$ is another strategy profile such that $f_i''$ is either $f_i$ or $f_i'$ for every $i$, then $f''$ is an $\ve$-Nash equilibrium that satisfies
\[\max_{i\in \{1,...,n\}}|U_i(f'')-U_i(f)|<\ve.\]
\end{enumerate}
Note that, by 2), $f'$ itself is also an $\ve$-Nash equilibrium.

Then, the following result is obtained.

\vspace{12pt}
\noindent
{\bf Theorem 3}. If the game is usual and $\mu$ is weakly conditionally atomless for all players, then for every $\ve>0$ and every Nash equilibrium $f$, there exists an $\ve$-purification $f'$ of $f$ such that, for each player $i$, the range of $f_i'$ is finite.

\vspace{12pt}
Combining our Theorem 3 and Balder's (1988) main theorem, we obtain the following result. Let $\mu_i$ be the marginal probability of $\mu$ in the space $X_i$.

\vspace{12pt}
\noindent
{\bf Theorem 4}. Suppose that the game is usual, $\mu$ is weakly conditionally atomless for all players, and $\mu$ is absolutely continuous with respect to $\mu_1\otimes\cdots\otimes\mu_n$. Then, for every $\ve>0$, there exists an $\ve$-Nash equilibrium such that each player $i$ chooses a pure strategy $f_i$ and the range of $f_i$ is finite.

\section{Discussion}
In related research, our definition of equivalence is sometimes called the {\bf payoff equivalence}. There is another notion of equivalence, called the {\bf distributional equivalence}. Two strategies $f_i$ and $f_i'$ are distributionally equivalent if, for every Borel measurable set $\Lambda\subset K_i$,
\[\int 1_{\{k_i\in \Lambda\}}df_{i,x_i}d\mu=\int 1_{\{k_i\in \Lambda\}}df_{i,x_i}'d\mu.\]
This definition is on the exact purification. For the approximated purification, we can define the distributional equivalence as follows: two strategy $f_i$ and $f_i'$ are $\ve$-distributionally equivalent if, for every Borel measurable set $\Lambda\subset K_i$,
\[\left|\int 1_{\{k_i\in \Lambda\}}d(f_{i,x_i}-f_{i,x_i}')d\mu\right|<\ve.\]
If $K_i$ is finite, then the number of possible $\Lambda$ is also finite. Thus, if once we obtain a purification result for the payoff equivalence in the setup of this paper, then we can immediately obtain a purification result for the distributional equivalence in the following manner: if the possibility of $\Lambda$ is $\Lambda_1,...,\Lambda_L$, then for $\Lambda_{\ell}$, define
\[u_{m+\ell}(k,x)=\begin{cases}
1 & \mbox{if }k_i\in \Lambda_{\ell},\\
0 & \mbox{otherwise}.
\end{cases}\]
It is obvious that a result for the payoff equivalence for this $u$ leads to a result for the distributional equivalence. However, if $K_i$ is infinite, then the above method cannot be used, because the range of $u$ becomes infinite dimensional.

In many previous studies, the finiteness of $K_i$ and some independence of the signal structure were assumed, and then the existence of an exact purification in both the payoff and distributional senses was proved (see Milgrom and Weber (1985), Radner and Rosenthal (1982), Khan et al. (2006)). The finiteness requirement of $K_i$ can be replaced with a countability requirement (see Khan and Rath (2009)). However, if $K_i$ is uncountable, there is an impossibility theorem regarding exact purification. That is, if $K_i$ is uncountable, then there may be no exact purification, even though the signal structure satisfies the usual independence requirement (see Khan and Zhang (2014)).

Our Theorem 4 is an existence theorem for an approximated Nash equilibrium with pure strategies. The basis for this result is an existence theorem for Nash equilibria with mixed strategies by Balder (1988). For this result, we use the absolute continuity of $\mu$ with respect to $\mu_1\otimes ... \otimes \mu_n$, and independence assumption is not needed. Similar results have already been produced for a correlated equilibrium (see Cotter (1991) and Stinchcombe (2011)). Based on these results, one might think that not only the approximated purification, but also the exact purification can be obtained from the absolute continuity assumption alone, and the independence assumption can be avoided. However, a counterexample for this conjecture has already been obtained in Example 1 of Yu et al. (2018). Note that, in this counterexample, all assumptions in Theorem 4 hold. Therefore, independence is crucial for the exact purification.

We should mention the notion of saturation. Although there are many equivalent definitions of saturation, we think that the definition using the ``essentially countably generated'' assumption is relatively easy to understand. For a given finite measure space $(\Omega,\mathscr{S},\mu)$, define $\mathscr{N}=\{N\in \mathscr{S}|\mu(N)=0\}$. This space is said to be {\bf essentially countably generated} if there exists a countable family $\mathscr{C}\subset \mathscr{S}$ such that the smallest $\sigma$-algebra that contains $\mathscr{C}\cup\mathscr{N}$ coincides with $\mathscr{S}$.\footnote{This definition is different from that in Khan and Zhang (2014). However, we can easily check that these two definitions coincide, by almost the same arguments as in the proof of Proposition 3.3.2 in Dudley (2002).} Now, choose any $A\in \mathscr{S}$ such that $\mu(A)>0$ and define $\mathscr{S}_A=\{B\in\mathscr{S}|B\subset A\}$. Then, $(A,\mathscr{S}_A,\mu)$ is also a finite measure space. The finite measure space $(\Omega,\mathscr{S},\mu)$ is said to be {\bf saturated} if and only if there is no $A\in \mathscr{S}$ such that $\mu(A)>0$ and $(A,\mathscr{S}_A,\mu)$ is essentially countably generated.

Keisler and Sun (2009) showed that if a space of infinitely many players and an uncountable action set are given, then this space is saturated if and only if every game with this player set and action set has a Nash equilibrium. A similar result was derived by Khan and Sagara (2016) for a Walrasian equilibrium of an economy such that there are infinitely many agents and the commodity space is included in $L^{\infty}$. To the best of our understanding, the above results require saturation because the existence of such an equilibrium is deeply related to Lyapunov's convexity theorem for Bochner multi-valued integrals in infinite dimensional spaces, which is known as an equivalent condition for saturation.

For the theory of exact purifications, the Dvoretzky-Wald-Wolfowitz theorem is crucial, and the proof of this theorem requires Lyapunov's convexity theorem. Therefore, if the action set is uncountable, then we are confronted with the problem that Lyapunov's convexity theorem cannot be applied when the space is not saturated. Theorem 2 of Khan and Zhang (2014) is one of the straightforward consequences of this fact, which states that if the signal structure is not saturated, we can construct a game in which there is no pure strategy Nash equilibrium. A concrete example of such a game was obtained by Khan et al. (1999).

In contrast, for the theory of approximate purifications, Lyapunov's convexity theorem is not crucial. This is the main reason why our theorems hold for possibly non-saturated signal structures. Note that, in many applied research, the signal structure is assumed to be a Polish space with a Borel probability measure, which is never saturated because every Polish space with a Borel probability measure is essentially countably generated.\footnote{Note that every Polish space is second-countable.} Therefore, we think that our results are worthwhile.

\section{Conclusion}
We showed that for only weak atomless assumption on the prior, there exists an approximate purification for any mixed strategy. This atomless requirement could be further weakened when considering a purification of a Nash equilibrium. We did not need any independence assumption on signal structure to show these results. Using these results, we have succeeded in obtaining an existence theorem for an $\ve$-Nash equilibrium under sufficiently weak assumptions.

All of these results were obtained under the assumption that all action sets are compact. In most of the previous studies, action sets are assumed to be finite, and in this sense, our results are a generalization of these results. Since it has been shown that this generalization is not possible in the context of exact purification, we can say that our result is a non-trivial generalization.

It is not known whether it is possible to remove compactness of the action sets. Since we have already used the compactness of the space in Step 1 of the proof of Theorem 1, it is completely unknown whether such a generalization is possible.

\section{Proofs}
\subsection{Proof of Lemma 1}
Because $\mathscr{P}$ is countable, we can set $\mathscr{P}=\{P_1,P_2,...\}$. Note that, because the function $P\mapsto \rho(P_i,P)$ is continuous with respect to $\rho$, we have that $v_i:x\mapsto \rho(P_i,f(x))$ is measurable. For each $\nu$, define $I_{\nu}(x)=\{i\in \{1,...,\nu\}|v_i(x)=\min_{j\in \{1,...,\nu\}}v_j(x)\}$, and $f_{\nu}(x)=P_i$ for $i=\min I_{\nu}(x)$. Then,
\begin{align*}
f_{\nu}(x)=P_i\Leftrightarrow&~v_i(x)<v_j(x)\mbox{ for every }j\in \{1,...,i-1\}\\
&\mbox{ and }v_i(x)\le v_j(x)\mbox{ for every }j\in \{i+1,...,\nu\}.
\end{align*}
Thus, the function $f_{\nu}$ is simple and for every $x$, $f_{\nu}(x)\to f(x)$ as $\nu\to \infty$, which completes the proof. $\blacksquare$

\subsection{Proof of Lemma 2}
Suppose that $f:X\to \mathscr{M}^1_+(K)$ is a function such that for every Borel set $B\subset K$, $x\mapsto f_x(B)$ is measurable. We will show that this function is weakly measurable.

Since $\mathscr{M}_+^1(K)$ is compact, there exists a countable dense subset $\mathscr{P}=\{P_1,P_2,...\}$ of $\mathscr{M}_+^1(K)$. For each $i$, define
\[v_i(x)=\rho(f_x,P_i),\]
where $\rho$ is the Prohorov metric of $\mathscr{M}_+^1(K)$. We show that each $v_i$ is measurable. Recall that for $\ve>0$ and $A\subset K$,
\[A^{\ve}=\{k\in K|\exists k'\in K\mbox{ s.t. }d(k,k')<\ve\},\]
where $d$ is the metric of $K$. For each Borel set $A\subset K$, define
\[\eta_A(x)=\inf\{\ve>0|f_x(A)\le P_i(A^{\ve})+\ve\}.\]
Then, by the definition of the Prohorov metric,
\[v_i(x)=\sup_{C\in\mathscr{C}}\eta_C(x),\]
where $\mathscr{C}$ denotes the set of all closed subsets of $K$.\footnote{See section 11.3 of Dudley (2002).} Now, for each $C\in\mathscr{C}$ and each rational number $q\in [0,1]$, define
\[\varphi_{C,q}(x)=\begin{cases}
q & \mbox{if }f_x(C)\le P_i(C^q)+q, \\
1 & \mbox{otherwise}.
\end{cases}\]
Since $x \mapsto f_x(C)$ is measurable, $\varphi_{C,q}$ is measurable. Therefore, for each $C\in\mathscr{C}$, $\eta_C=\inf_{q\in [0,1]\cap\mathbb{Q}}\varphi_{C,q}$ is measurable.

Because $K$ is compact, it is second countable, and thus it has a countable basis $\mathscr{V}=\{V_1,V_2,...\}$. Define
\[\tilde{\mathscr{C}}=\{C\in\mathscr{C}|C=(\cup_{j\in M}V_j)^c\mbox{ for some finite set }M\subset \mathbb{N}\}.\]
Since $\mathscr{V}$ is countable, so is $\tilde{\mathscr{C}}$. Fix $x\in X$ and $C\in\mathscr{C}$. There exists a subfamily $\mathscr{V}'$ of $\mathscr{V}$ such that $C=(\cup_{V\in\mathscr{V}'}V)^c$, and thus, there exists a decreasing sequence $(A_j)_{j\in\mathbb{N}}$ of sets in $\tilde{\mathscr{C}}$ such that $A_j\downarrow C$ as $j\to \infty$. Let $\ve^*=\sup_{j\in \mathbb{N}}\eta_{A_j}(x)$ and choose an $\ve>0$. Then, by the definition of $\ve^*$,
\[f_x(A_j)\le P_i((A_j)^{\ve^*+\ve})+\ve^*+\ve,\]
for each $j\in \mathbb{N}$. Therefore, by taking the limit $j\to \infty$, we have
\[f_x(C)\le P_i(\cap_j(A_j)^{\ve^*+\ve})+\ve^*+\ve.\]
We show that $\cap_{j=1}^{\infty}(A_j)^{\ve^*+\ve}\subset C^{\ve^*+2\ve}$. Choose any $k\in\cap_j(A_j)^{\ve^*+\ve}$. Then, for each $j$, there exists $k^j \in A_j$ such that $d(k,k^j)<\ve^*+\ve$. Since $K$ is compact, we have that there is a limit point $k^*$ of the sequence $(k^j)$. By the definition of $A_j$, we have that $k^*\in C$. Because $d(k,k^*)\le\ve^*+\ve<\ve^*+2\ve$, we have that $k\in C^{\ve^*+2\ve}$, as desired. Therefore,
\[f_x(C) \le P_i(C^{\ve^*+2\ve})+\ve^*+2\ve.\]
Since $\ve>0$ is arbitrary, it follows that $\eta_C(x)\le\ve^*$. Therefore, for each $x\in X$,
\[v_i(x)=\sup_{C\in\mathscr{C}}\eta_C(x)=\sup_{C\in\tilde{\mathscr{C}}}\eta_C(x),\]
and thus $v_i$ is measurable.

Now, for each $\nu$, let $I_{\nu}(x)=\{i\in \{1,...,\nu\}|v_i(x)=\min_{j\in \{1,..,\nu\}}v_j(x)\}$ for each $x\in X$ and define $f_x^{\nu}=P_i$ for $i=\min I_{\nu}(x)$. Then,
\begin{align*}
f_x^{\nu}=P_i\Leftrightarrow&~v_i(x)<v_j(x)\mbox{ for every }j\in \{1,...,i-1\}\\
&\mbox{ and }v_i(x)\le v_j(x)\mbox{ for every }j\in \{i+1,...,\nu\}.
\end{align*}
Because each $v_j$ is measurable, we have that $f^{\nu}$ is a simple function. Moreover, $\rho(f_x^{\nu},f_x)\to 0$ as $\nu\to\infty$. Let $\tau:\mathscr{M}^1_+(K)\to \mathbb{R}$ be a continuous function. Then, for each $\nu$, $\tau\circ f^{\nu}$ is a simple function. Moreover, $\tau\circ f^{\nu}\to\tau\circ f$ pointwise as $\nu\to\infty$, and hence $\tau\circ f$ is measurable. This implies that $f$ is weakly measurable, as desired.

Next, we show the opposite direction. Let $f:X\to\mathscr{M}_+^1(K)$ be a weakly measurable function. It suffices to show that for every Borel set $B$ of $K$, the following function $\zeta_B:x\mapsto f_x(B)$ is measurable. First, for each continuous function $\psi:K\to \mathbb{R}$,
\[\xi_{\psi}:Q\mapsto\int\psi dQ\]
is continuous. Because $f$ is weakly measurable, for each continuous function $\psi:K\to\mathbb{R}$, $\xi_{\psi}\circ f$ is measurable. Next, let $\mathscr{L}$ be the family of Borel sets $B\subset K$ such that $\zeta_B$ is measurable on $X$. Fix a closed set $C\subset K$. For each $\nu$, let
\[\psi_{\nu}(k)=\max\{1-\nu\inf_{k' \in C}d(k,k'),0\}.\]
Then, $\psi_{\nu}$ is continuous. For each $k\in K$, $\psi_{\nu}(k)\downarrow 1_C(k)$ as $\nu\to \infty$, and by the monotone convergence theorem, we have that for each $x\in X$, $\int\psi_{\nu}df_x\to\int 1_Cdf_x$. Therefore, the sequence $(\xi_{\psi_{\nu}}\circ f)$ converges pointwise to the function $\zeta_C$ as $\nu\to \infty$, which implies that $C\in \mathscr{L}$.

Since $\zeta_{K\setminus B}=1-\zeta_B$, we have that $B\in\mathscr{L}$ implies $K\setminus B\in\mathscr{L}$. Therefore, $\mathscr{L}$ contains all open sets. Moreover, if $B_1\in\mathscr{L}$ is open and $B_2\in\mathscr{L}$ is closed, then $B_1\cap B_2\in\mathscr{L}$ because $\zeta_{B_1\cap B_2}=\zeta_{B_1}-\zeta_{B_1\setminus B_2}$ is measurable. Let $\mathscr{D}$ be the set of all Borel sets such that there exists an open $B_1$ and a closed $B_2$ such that $B=B_1\cap B_2$. Then, $\mathscr{D}$ is a semiring and $\mathscr{D}\subset \mathscr{L}$.\footnote{A family $\mathscr{D}$ of subsets of $Z$ is called a {\it semiring} if 1) $\emptyset\in\mathscr{D}$, 2) for each pair $A,B\in\mathscr{D}$, $A\cap B \in \mathscr{D}$, and 3) for each pair $A,B\in\mathscr{D}$, $A\setminus B=\bigcup_{j=1}^kC_j$ for some $k$ and disjoint $C_1,...,C_k\in\mathscr{D}$. See section 3.2 of Dudley (2002) for detailed arguments.} Next, define
\[\mathscr{A}=\{B|B=\cup_{j=1}^{\nu}D_j\mbox{ for some }\nu\mbox{ and disjoint }D_1,...,D_{\nu}\in\mathscr{D}\}.\]
Then, $\mathscr{A}$ is an algebra that includes all open sets.\footnote{See Proposition 3.2.3 of Dudley (2002).} Note that, for each disjoint pair $B_1,B_2\in\mathscr{L}$, we have $B_1\cup B_2\in\mathscr{L}$ because $\zeta_{B_1\cup B_2}=\zeta_{B_1}+\zeta_{B_2}$. This implies that $\mathscr{A}\subset\mathscr{L}$. If $(A_{\nu})$ is an increasing (resp. decreasing) sequence of sets in $\mathscr{L}$ with $A_{\nu}\uparrow A$ (resp. $A_{\nu}\downarrow A$) as $\nu\to\infty$, then by the monotone convergence theorem, we have that $\zeta_{A_{\nu}}\to\zeta_A$ pointwise, and thus $A\in\mathscr{L}$. This implies that $\mathscr{L}$ is a monotone class that contains $\mathscr{A}$. By the monotone class lemma,\footnote{See Theorem 4.4.2 of Dudley (2002).} we have that $\mathscr{L}$ is the same as the Borel $\sigma$-field of $K$. Thus, $\zeta_B$ is measurable for any Borel set $B\subset K$. This completes the proof. $\blacksquare$

\subsection{Proof of Theorem 1}
Throughout this proof, we treat $\rho$ as the Prohorov metric of either $\mathscr{M}^1_+(K)$ or $\mathscr{M}^1_+(L)$. We consider that the abbreviation of the notation $K$ or $L$ for Prohorov metrics should not cause any confusion.

Note that, because $K$ is compact, there exists a countable dense family $\mathscr{P}=\{P_1,P_2,...\}$ of $\mathscr{M}^1_+(K)$ such that each $P_i$ has a finite support. We call a mixed strategy {\bf simple} if it is a simple function as a function from $X$ into $\mathscr{M}^1_+(K)$.

We separate the proof into eight steps.

\vspace{12pt}
\noindent
{\bf Step 1}. Suppose that $f$ is a mixed strategy of player $1$, and $(f_{\nu})$ is a sequence of simple mixed strategies of player $1$ that converges to $f$ pointwise as $\nu\to \infty$. Then, for every $(x,y)\in X\times Y$ and $i\in \{1,...,m\}$, the following holds.
\[\lim_{\nu\to \infty}\sup_{\ell\in L}\left|\int_Ku_i(k,\ell,x,y)d(f_x-f_{\nu,x})\right|=0.\]

\vspace{12pt}
\noindent
{\bf Proof of Step 1}. Suppose not. By taking a subsequence, we can assume that there exists $\delta>0$ and a sequence $(\ell_{\nu})$ of $L$ such that
\[\left|\int_Ku_i(k,\ell_{\nu},x,y)d(f_x-f_{\nu,x})\right|\ge \delta.\]
Because $L$ is compact, we can assume without loss of generality that $\ell_{\nu}\to \ell^*\in L$ as $\nu\to \infty$. Moreover, because $K$ and $L$ are compact, the function $(k,\ell)\mapsto u_i(k,\ell,x,y)$ is uniformly continuous, and thus for the metrics $d_K$ of $K$ and $d_L$ of $L$, there exists $\delta'>0$ such that if $d_K(k,k')+d_L(\ell,\ell')<\delta'$, then $|u_i(k,\ell,x,y)-u_i(k',\ell',x,y)|<\frac{\delta}{8}$. Because $f_{\nu,x}$ converges to $f_x$ with respect to the weak* topology, for sufficiently large $\nu$, $d_L(\ell_{\nu},\ell^*)<\delta'$ and
\[\left|\int_Ku_i(k,\ell^*,x,y)d(f_{\nu,x}-f_x)\right|<\frac{\delta}{2}.\]
Therefore, for any such $\nu$,
\begin{align*}
&~\int_Ku_i(k,\ell_{\nu},x,y)d(f_{\nu,x}-f_x)\\
\le&~\int_K\left(u_i(k,\ell^*,x,y)+\frac{\delta}{8}\right)d(f_{\nu,x}-f_x)^+\\
&~-\int_K\left(u_i(k,\ell^*,x,y)-\frac{\delta}{8}\right)d(f_{\nu,x}-f_x)^-\\
\le&~\frac{\delta}{2}+\int_Ku_i(k,\ell^*,x,y)d(f_{\nu,x}-f_x)<\delta,
\end{align*}
where $(f_{\nu,x}-f_x)^+$ (resp. $(f_{\nu,x}-f_x)^-$) is the positive (resp. negative) part of the Jordan decomposition of the measure $f_{\nu,x}-f_x$. By symmetric arguments, we can show that
\[\int_Ku_i(k,\ell_{\nu},x,y)d(f_{\nu,x}-f_x)>-\delta,\]
and thus,
\[\left|\int_Ku_i(k,\ell_{\nu},x,y)d(f_{\nu,x}-f_x)\right|<\delta,\]
which is a contradiction. This completes the proof of Step 1. $\blacksquare$

\vspace{12pt}
Fix any $\ve>0$ and any mixed strategy $f$ of player $1$. By Lemma 1, there exists a sequence $(f_{\nu})$ of simple mixed strategies of player $1$ such that $f_{\nu}\to f$ pointwise as $\nu\to \infty$ and the range of $f_{\nu}$ is always included in $\mathscr{P}$. By Step 1 and the dominated convergence theorem, there exists $\nu$ such that
\[\int_{X\times Y}\sup_{\ell\in L}\left|\int_Ku_i(k,\ell,x,y)d(f_x-f_{\nu,x})\right|d\mu<\frac{\ve}{2}\]
for every $i\in \{1,...,m\}$. Set $f''=f_{\nu}$, and let $K'$ be the union of the support of $f_x''$. Then, $K'$ is a finite set. Moreover, for every mixed strategy $g$ of player 2,
\begin{align*}
|U_i(f'',g)-U_i(f,g)|=&~\left|\int_{X\times Y}\int_{K\times L}u_i(k,\ell,x,y)d((f_x''-f_x)\otimes g_y)d\mu\right|\\
\le&~\int_{X\times Y}\int_L\left|\int_Ku_i(k,\ell,x,y)d(f_x''-f_x)\right|dg_yd\mu\\
\le&~\int_{X\times Y}\sup_{\ell\in L}\left|\int_Ku_i(k,\ell,x,y)d(f_x''-f_x)\right|d\mu\\
<&~\frac{\ve}{2}.
\end{align*}

\vspace{12pt}
\noindent
{\bf Step 2}. There exists $\delta>0$ such that for any two strategies $g^1$ and $g^2$ of player $2$ and any strategy $f'$ of player $1$, if $\rho(g^1_y,g^2_y)<\delta$ for every $y\in Y$ and the support of $f_x'$ is included in $K'$ for every $x\in X$, then for every $i\in \{1,...,m\}$,
\[|U_i(f',g^1)-U_i(f',g^2)|<\frac{\ve}{6}.\]

\vspace{12pt}
\noindent
{\bf Proof of Step 2}. Suppose not. Then, for every positive integer $\nu$, we can choose $i, f_{\nu}, g^1_{\nu}, g^2_{\nu}$ such that the support of $f_{\nu,x}$ is included in $K'$ for every $x\in X$, $\rho(g^1_{\nu,y},g^2_{\nu,y})<\frac{1}{\nu}$ for every $y\in Y$, and
\[U_i(f_{\nu},g^1_{\nu})-U_i(f_{\nu},g^2_{\nu})\ge \frac{\ve}{6}.\]
Because $\mathscr{M}^1_+(L)$ is compact and $K'$ is finite, we have that the following mapping
\[P\mapsto \max_{k\in K'}\int_Lu_i(k,\ell,x,y)dP\]
is uniformly continuous with respect to the Prohorov metric $\rho$. Therefore,
\[\max_{k\in K'}\int_Lu_i(k,\ell,x,y)d(g^1_{\nu,y}-g^2_{\nu,y})\to 0\]
as $\nu\to \infty$. By the dominated convergence theorem,
\begin{align*}
\frac{\ve}{6}\le&~U_i(f_{\nu},g^1_{\nu})-U_i(f_{\nu},g^2_{\nu})\\
=&~\int_{X\times Y}\sum_{k\in K'}f_{\nu,x}(\{k\})\int_Lu_i(k,\ell,x,y)d(g^1_{\nu,y}-g^2_{\nu,y})d\mu\\
\le&~\int_{X\times Y}\max_{k\in K'}\int_Lu_i(k,\ell,x,y)d(g^1_{\nu,y}-g^2_{\nu,y})d\mu\\
\to&~0\mbox{ as }\nu\to \infty,
\end{align*}
which is a contradiction. This completes the proof of Step 2. $\blacksquare$

\vspace{12pt}
\noindent
{\bf Step 3}. There exist a finite subset $L'\subset L$ and $\{Q_1,...,Q_N\}\subset M^1_+(L)$ such that the support of $Q_j$ is included in $L'$ for all $j$, and for every mixed strategy $g$ of player $2$, there exists a simple mixed strategy $g'$ of player $2$ such that $g_y'=Q_j$ for some $j$ and $\rho(g_y,g_y')<\delta$ for every $y\in Y$, where $\delta>0$ is given in Step 2.

\vspace{12pt}
\noindent
{\bf Proof of Step 3}. Because $L$ is compact, there exists a countable dense subset $\mathscr{Q}$ of $\mathscr{M}^1_+(L)$ such that every $Q\in \mathscr{Q}$ has a finite support. Therefore, there exists a finite set $L'\subset L$ and $\{Q_1,...,Q_N\}\subset \mathscr{M}^1_+(L)$ such that the support of $Q_j$ is included in $L'$ for all $j$, and for every $Q\in \mathscr{M}^1_+(L)$, there exists $j$ such that $\rho(Q,Q_j)<\delta$. Choose any mixed strategy $g$ of player $2$, and define $g_y'=Q_{j^*}$, where $j^*=\min\{j|\rho(g_y,Q_j)<\delta\}$. Then, $g'$ is a simple mixed strategy such that $g_y'=Q_j$ for some $j$ and $\rho(g_y,g_y')<\delta$ for every $y\in Y$, as desired. This completes the proof of Step 3. $\blacksquare$

\vspace{12pt}
Until the end of Step 7, we assume that $u_i$ is nonnegative for all $i$. Define
\[c=\sum_{(i,k,\ell)\in \{1,...,m\}\times K'\times L'}\int_{X\times Y}u_i(k,\ell,x,y)d\mu.\]
If $c=0$, then we can add $u_{m+1}(k,l,x,y)\equiv 1$, and thus we can assume without loss of generality that $c>0$. Define $Y'=Y\times \{1,...,m\}\times K'\times L'$ and a measure $\nu$ on the set $X\times Y'$ such that
\[\nu(A\times \{(i,k,\ell)\})=c^{-1}\int_Au_i(k,\ell,x,y)d\mu.\]
Define
\[c_{i,k,\ell}(y)=\int_Xu_i(k,\ell,x,y)\mu(dx|y),\]
\[d_{i,k,\ell}(y)=\begin{cases}
0 & \mbox{if }c_{i,k,\ell}(y)=0,\\
\frac{1}{c_{i,k,\ell}(y)} & \mbox{otherwise}.
\end{cases}\]
If $y'=(y,i,k,\ell)$, then
\[\nu(T|y')=d_{i,k,\ell}(y)\int_Tu_i(k,\ell,x,y)\mu(dx|y),\]
\[\nu_{Y'}(S\times \{(i,k,\ell)\})=c^{-1}\int_Sc_{i,k,\ell}(y)d\mu_Y.\]
In particular, $d_{i,k,\ell}(y)=\frac{1}{c_{i,k,\ell}(y)}$ for almost all $(y,i,k,\ell)\in Y'$ with respect to $\nu_{Y'}$, and thus $\nu$ is conditionally atomless for player $1$.

\vspace{12pt}
\noindent
{\bf Step 4}. Let $T$ be a measurable subset of $X$. Then, there exists a sequence of the partitions $(\{H_1^M,...,H_M^M\})$ of $T$ such that 
\[\lim_{M\to \infty}\int_{Y'}\left(\max_j\nu(H_j^M|y')\right)d\nu_{Y'}=0.\]

\vspace{12pt}
\noindent
{\bf Proof of Step 4}. By assumption on $X$, we can assume without loss of generality that $X=[0,1]$, and the $\sigma$-algebra $\mathscr{F}$ of $X$ includes all Borel sets. First, we show that every probability measure $P$ on $([0,1],\mathscr{F})$ is atomless if and only if the cumulative distribution function $F(x)=P([0,x])$ satisfies $F(0)=0$ and is uniformly continuous.

Suppose that $P$ is atomless. Then, $P(\{x\})=0$ for all $x\in [0,1]$, and thus $F(0)=0$ and $F$ is left-continuous. Because every cumulative distribution function is right-continuous, we have that $F$ is continuous. Since $[0,1]$ is compact, $F$ is uniformly continuous.

Conversely, suppose that $F(0)=0$ and $F$ is uniformly continuous. Then, $P(\{x\})=0$ for all $x\in [0,1]$. For every set $B\in\mathscr{F}$,
\begin{align*}
&~\max_{i\in \{1,...,m\}}P(B\cap [(i-1)/m,i/m])\\
\le&~\max_{i\in \{1,...,m\}}[F(i/m)-F((i-1)/m)]\to 0\mbox{ as }m\to \infty,
\end{align*}
which implies that $B$ is not an atom of $P$. Thus, our claim is correct.

For $M\in \mathbb{N}$, define
\[H_1^M=T\cap [0,M^{-1}],\]
and for $j\in \{2,...,M\}$,
\[H_j^M=T\cap ](j-1)M^{-1},jM^{-1}].\]
We show that this sequence $(\{H_1^M,...,H_M^M\})$ of partitions of $T$ satisfies the requirement of our claim.

Fix any $y'\in Y'$ such that $\nu(\cdot|y')$ is atomless. By our previous argument, we have that the cumulative distribution function $x\mapsto \nu([0,x]|y')$ is uniformly continuous and $\nu(\{0\}|y')=0$, and thus
\[\lim_{M\to \infty}\left(\max_j\nu(H_j^M|y')\right)=0.\]
Because $\nu$ is conditionally atomless for player $1$, we have that $\nu(\cdot|y')$ is atomless for almost all $y'$ with respect to $\nu_{Y'}$. Therefore, by the dominated convergence theorem, we have that
\[\lim_{M\to \infty}\int_{Y'}\left(\max_j\nu(H_j^M|y')\right)d\nu_{Y'}=0,\]
as desired. This completes the proof of Step 4. $\blacksquare$

\vspace{12pt}
Let $\mathbb{R}^{K'}$ be the set of all real-valued function on $K'$. For $s\in \mathbb{R}^{K'}$, we write $s_k$ instead of $s(k)$, and define $\|s\|=\sqrt{\sum_{k\in K'}s_k^2}$ as usual. Let $\Delta^{K'}$ be the set of all $s\in \mathbb{R}^{K'}$ such that $s_k\ge 0$ for all $k\in K'$ and $\sum_{k\in K'}s_k=1$. Moreover, let $V^{K'}$ be the set of all $s\in \mathbb{R}^{K'}$ such that there exists $k\in K'$ such that $s_k=1$ and $s_{k'}=0$ for all $k'\in K'\setminus \{k\}$.

\vspace{12pt}
\noindent
{\bf Step 5}. Suppose that $s\in \Delta^{K'}$. Then, for every measurable set $T$ in $X$ and $\kappa>0$, there exists a measurable function $b:T\to V^{K'}$ such that
\[\int_{Y'}\left\|\nu(T|y')s-\int_Tb(x)\nu(dx|y')\right\|d\nu_{Y'}<\kappa.\]

\vspace{12pt}
\noindent
{\bf Proof of Step 5}. Let $\Omega'$ be some probability measure space and $Z^1,Z^2,...$ be an independent family of random variables defined on $\Omega'$ such that $Z^j(\omega)\in V^{K'}$ and $E(Z^j)=s$.\footnote{The existence of such $\Omega'$ and $Z^1,Z^2,...$ can easily be shown. For example, let $\Omega_i$ be a copy of $K'$ with probability $P_i$ such that $P_i(\{k\})=s_k$. Let $\Omega'$ be the product probability space $\prod_{i=1}^{\infty}\Omega_i$, and
\[Z^i_{k}(k_1,k_2,...)=\begin{cases}
1 & \mbox{if }k=k_i,\\
0 & \mbox{if }k\neq k_i.
\end{cases}\]
Then, $Z^1,Z^2,...$ satisfies all our requirements.} Note that, by definition, we have that $\sum_k\mbox{Var}(Z_k^j)\le 1$. Define $H_1^M,...,H_M^M$ as in Step 4, and
\[b_M(x,\omega)=Z^j(\omega)\mbox{ if }x\in H_j^M.\]
Then, we have that
\[\int_Tb_M(x,\omega)\nu(dx|y')=\sum_{j=1}^M\nu(H_j^M|y)Z^j(\omega),\]
and thus,
\[E\left(\int_Tb_M(x,\omega)\nu(dx|y')\right)=\nu(T|y')s.\]
Meanwhile, for almost all $y'\in Y'$,
\begin{align*}
&~E\left(\left\|\nu(T|y')s-\int_Tb_M(x,\omega)\nu(dx|y')\right\|^2\right)\\
=&~E\left(\left\|\nu(T|y')s-\sum_j\nu(H_j^M|y')Z^j(\omega)\right\|^2\right)\\
=&~\sum_k\mbox{Var}\left(\sum_j\nu(H_j^M|y')Z_k^j\right)\\
=&~\sum_j(\nu(H_j^M|y'))^2\sum_k\mbox{Var}(Z_k^j)\\
\le&~\sum_j(\nu(H_j^M|y'))^2\\
\le&~\max_j\nu(H_j^M|y'),
\end{align*}
where the last inequality follows from H\"older's inequality. Therefore, by Step 4 and Fubini's theorem, for sufficiently large $M$, we have that
\begin{align*}
&~E\left(\int_{Y'}\left\|\nu(T|y')s-\int_Tb_M(x,\omega)\nu(dx|y')\right\|^2d\nu_{Y'}\right)\\
=&~\int_{Y'}E\left(\left\|\nu(T|y')s-\int_Tb_M(x,\omega)\nu(dx|y')\right\|^2\right)d\nu_{Y'}<\kappa^2.
\end{align*}
Hence, there exists $\omega\in \Omega'$ such that
\begin{align*}
&~\left(\int_{Y'}\left\|\nu(T|y')s-\int_Tb_M(x,\omega)\nu(dx|y')\right\|d\nu_{Y'}\right)^2\\
\le&~\int_{Y'}\left\|\nu(T|y')s-\int_Tb_M(x,\omega)\nu(dx|y')\right\|^2d\nu_{Y'}<\kappa^2,
\end{align*}
where the first inequality is verified by the Cauchy-Schwarz inequality. Hence, we can set $b(x)=b_M(x,\omega)$. This completes the proof of Step 5. $\blacksquare$

\vspace{12pt}
For every bounded measurable function $h:X\to \mathbb{R}^{K'}$, define a seminorm
\[\|h\|_{\nu}=\int_{Y'}\left\|\int_Xh(x)\nu(dx|y')\right\|d\nu_{Y'}.\]
Note that, if $h$ is a mixed strategy of player $1$ and the support of $h_x$ is always included in $K'$, then $h_x$ can be seen as an element of $\Delta^{K'}$, and thus $h$ can be treated as a bounded measurable function from $X$ into $\mathbb{R}^{K'}$, and the seminorm $\|h\|_{\nu}$ can be defined.

\vspace{12pt}
\noindent
{\bf Step 6}. For every $\kappa>0$ and every mixed strategy $h$ of player $1$ such that $h_x\in \Delta^{K'}$ for all $x\in X$, there exists a pure strategy $h'$ of player 1 such that $h_x'\in \Delta^{K'}$ for all $x\in X$ and 
\[\|h-h'\|_{\nu}<\kappa.\]

\vspace{12pt}
\noindent
{\bf Proof of Step 6}. Clearly, $h:X\to \Delta^{K'}$ is measurable in the usual sense, and thus, by the dominated convergence theorem, there exists a simple mixed strategy $h'':X\to \Delta^{K'}$ such that
\[\|h-h''\|_{\nu}\le \frac{\kappa}{2}.\]
Meanwhile, the range of $h''$ is a finite set $\{p^1,...,p^q\}$. Let $T_j=(h'')^{-1}(p^j)$. Then, applying Step 5 for $T=T_j$ and $s=p^j$, we have that there exists a measurable function $b^j:T_j\to V^{K'}$ such that
\[\int_{Y'}\left\|\int_{T_j}(h_x''-b^j(x))\nu(dx|y')\right\|d\nu_{Y'}<\frac{\kappa}{2q}.\]
Define $h_x'=b^j(x)$ if $x\in T_j$. Then, $h'$ is a function from $X$ into $V^{K'}$, and thus it is actually a pure strategy. Moreover,
\[\|h''-h'\|_{\nu}=\int_{Y'}\left\|\int_X(h_x''-h_x')\nu(dx|y')\right\|d\nu_{Y'}<\frac{\kappa}{2}.\]
Thus,
\[\|h-h'\|_{\nu}\le \|h-h''\|_{\nu}+\|h''-h'\|_{\nu}<\kappa,\]
as desired. This completes the proof of Step 6. $\blacksquare$

\vspace{12pt}
\noindent
{\bf Step 7}. The claim of this theorem is correct if $u_i$ is nonnegative for each $i$.

\vspace{12pt}
\noindent
{\bf Proof of Step 7}. Recall that we have already fixed $\ve$ and $f, f''$ before Step 2. By Step 6, there exists a pure strategy $f'$ such that
\[\|f''-f'\|_{\nu}<\frac{\ve}{6c|K'||L'|}.\]
Choose any strategy $g$ of player $2$. By Step 3, there exists a simple mixed strategy $g'$ of player $2$ such that the support of $g_y'$ is included in $L'$ and $\rho(g_y,g_y')<\delta$ for every $y\in Y$. For $h_k(x)=f_x''(\{k\})-f_x'(\{k\})$,
\begin{align*}
&~\left|\int_Yg_y'(\{\ell\})\int_Xh_k(x)u_i(k,\ell,x,y)\mu(dx|y)d\mu_Y\right|\\
\le&~\int_Yg_y'(\{\ell\})\left|\int_Xh_k(x)u_i(k,\ell,x,y)\mu(dx|y)\right|d\mu_Y\\
=&~c\int_{Y\times\{(i,k,\ell)\}}g_y'(\{\ell\})d_{i,k,\ell}(y)\left|\int_Xc_{i,k,\ell}(y)h_k(x)\nu(dx|y')\right|d\nu_{Y'}\\
=&~c\int_{Y\times \{(i,k,\ell)\}}g_y'(\{\ell\})\left|\int_Xh_k(x)\nu(dx|y')\right|d\nu_{Y'}\\
\le&~c\|f''-f'\|_{\nu}<\frac{\ve}{6|K'||L'|}.
\end{align*}
Therefore, we have that
\[|U_i(f'',g')-U_i(f',g')|<\frac{\ve}{6}.\]
By Step 2, this implies that
\begin{align*}
|U_i(f'',g)-U_i(f',g)|\le&~|U_i(f'',g)-U_i(f'',g')|+|U_i(f'',g')-U_i(f',g')|\\
&~+|U_i(f',g')-U_i(f',g)|\\
<&~\frac{\ve}{2}.
\end{align*}
Therefore,
\[|U_i(f,g)-U_i(f',g)|\le |U_i(f,g)-U_i(f'',g)|+|U_i(f'',g)-U_i(f',g)|<\ve,\]
as desired. This completes the proof of Step 7. $\blacksquare$

\vspace{12pt}
\noindent
\textbf{Step 8}. The claim of this theorem is correct.

\vspace{12pt}
\noindent
\textbf{Proof of Step 8}. Define
\[v_i(k,l,x,y)=\begin{cases}
\max\{u_i(k,l,x,y),0\} & \mbox{if }1\le i\le m,\\
\max\{-u_{i-m}(k,l,x,y),0\} & \mbox{if }m+1\le i\le 2m,
\end{cases}.\]
Then, applying Step 7 for a nonnegative function $v:K\times L\times X\times Y\to \mathbb{R}^{2m}$, we obtain the desired result. This completes the proof of Step 8. $\blacksquare$

\subsection{Proof of Theorem 2}
We construct a modified game of the original game as follows. First, there are two players, one is $i$ and another is $0$. The observation spaces, the common prior, and the payoff function are the same as those in the original game. The action set and the observation function of player $i$ are the same as those of the original game, whereas the action set and the observation function of player $0$ are $\prod_{j\neq i}K_j$ and $x_{-i}=(x_1,...,x_{i-1},x_{i+1},...,x_n)$, respectively. Because the prior $\mu$ is conditionally atomless for player $i$ in the original game, it has the same property in this modified game. Therefore, by Theorem 1, for each $\ve>0$ and each mixed strategy $f_i$ of player $i$, there exists a pure strategy $f_i'$ such that for some finite subset $K_i'$ of $K_i$, $f_{i,x_i}'\in K_i'$ for any $x_i\in X_i$ and for each mixed strategy $g$ of player $0$ in the modified game,
\[\max_{j\in \{1,...,m\}}|U_j(f_i,g)-U_j(f_i',g)|<\ve.\]
Now, for each $j\neq i$, choose any mixed strategy $f_j$ of player $j$ in the original game. Then, the profile $f_{-i}=(f_1,...,f_{i-1},f_{i+1},...,f_n)$ is a mixed strategy of player $0$ in the modified game. Therefore, by the above inequality, we have
\[\max_{j\in\{1,...,m\}}|U_j(f_i,f_{-i})-U_j(f_i',f_{-i})|<\ve,\]
as desired. This completes the proof. $\blacksquare$

\subsection{Proofs of Theorems 3-4}
Theorem 3.1 of Balder (1988) showed that under the assumptions of Theorem 4, there exists a Nash equilibrium such that every strategy is, in his terminology, a `transition probability'. Lemma 2 said that our `mixed strategy' is equivalent to this `transition probability' of Balder (1988). Therefore, Theorem 3 implies Theorem 4, and it suffices to show Theorem 3.

We need a lemma.

\vspace{12pt}
\noindent
{\bf Lemma 3}. Suppose that $G^1,...,G^m$ are two player games, and for every game $G^j$, player $1$ has the same observation set $X_0$ and action set $K_0$. Moreover, suppose that for any $j$, the common prior $\mu^j$ in the game $G^j$ is conditionally atomless for player $1$. Then, for every $\ve>0$ and every mixed strategy $f$ of player $1$, there exists a pure strategy $f'$ of player $1$ such that $f'$ is $\ve$-equivalent to $f$ in every game $G^j$, and the range of $f'$ is finite.

\vspace{12pt}
\noindent
{\bf Proof}. We can assume that the payoff function $u^j$ of the game $G^j$ is a real-valued function: if not, then we can replace $G^j$ with games $G^j_1,...,G^j_M$, where $M$ is the dimension of the range of $u^j$, every $G^j_i$ has the same observation sets, action sets, and the common prior as $G^j$, and the payoff function of $G^j_i$ is $u^j_i$. Let $Y^j$ and $L^j$ be the observation set and the action set for player $2$ in $G^j$, respectively. We can assume without loss of generality that $Y^1,...,Y^m$ are disjoint. Define $Y=Y^1\cup ...\cup Y^m$, $L=L^1\times ... \times L^m$, and
\[\mu(B)=\sum_{j=1}^m\frac{1}{m}\mu^j(B\cap (X_0\times Y^j))\]
for each measurable set $B\subset X_0\times Y$. Moreover, for each $j$, define
\[u_j(k_0,\ell,x_0,y)=\begin{cases}
u^j(k_0,\ell^j,x_0,y) &\mbox{if }y\in Y^j,\\
0 & \mbox{otherwise},
\end{cases}\]
where $\ell=(\ell^1,...,\ell^m)$. Consider the game $G=(\{1,2\},(K,L),(X_0\times Y,\mu),u)$. We can easily check that $\mu$ is conditionally atomless for player $1$. By Theorem 1, there exists an $(\ve/m)$-purification $f'$ of $f$ in the game $G$ such that the range of $f'$ is finite. For each $j$, choose any mixed strategy $g^j$ of player $2$ in game $G^j$ and $y_j^*\in Y^j$. If $y\in Y^j$, define
\[g_y=g_{y_1^*}^1\otimes...\otimes g_{y_{j-1}^*}^{j-1}\otimes g_y^j\otimes g_{y_{j+1}^*}^{j+1}\otimes...\otimes g_{y_m^*}^m.\]
By Lemma 2, we can easily check that $g$ is a mixed strategy of player $2$ in the game $G$. Therefore,
\[\frac{\varepsilon}{m}>|U_j(f,g)-U_j(f',g)|=\frac{1}{m}|U^j(f,g^j)-U^j(f',g^j)|,\]
which completes the proof of Lemma 3. $\blacksquare$

\vspace{12pt}
Choose any Nash equilibrium $f$. We construct pure strategies $f_1',...,f_n'$ recursively. Let $1\le m\le n$ and suppose that for $i<m$, $f_i'$ is already defined. Define
\[H_m=\{h|h_j=f_j\mbox{ for }j\ge m,\ h_j=f_j\mbox{ or }h_j=f_j'\mbox{ for }j<m\}.\]
Note that, this set is finite, and in particular, $H_1=\{f\}$. For any $j\neq m$ and $h\in H_m$, we construct a two player game $G^{mjh}=(\{m,j\},(K_m,K_j),(X_m\times X_j,\mu_{mj}),u^{mjh})$ by defining $u^{mjh}(k_m,k_j,x_m,x_j)$ from the following integral:
\[\int_{\prod_{\ell\in N\setminus \{m,j\}}(X_{\ell}\times K_{\ell})}u(k,x)d\left(\bigotimes_{\ell\in N\setminus\{m,j\}}h_{\ell,x_{\ell}}\right)\mu(dx_{-mj}|x_m,x_j),\]
where $x_{-mj}=(x_1,...,x_{m-1},x_{m+1},...,x_{j-1},x_{j+1},...,x_n)$. Because $\mu$ is weakly conditionally atomless, the family $(G^{mjh})$ satisfies all requirements of Lemma 3, and thus there exists a pure strategy $f_m'$ of player $m$ such that
\[\max_{i\in N}|U^{mjh}_i(f_m,g_j)-U^{mjh}_i(f_m',g_j)|<\frac{\ve}{3^{n-m+1}}\]
for any $j\neq m$ and mixed strategy $g_j$ of player $j$ in game $G^{mjh}$, and the range of $f_m'$ is finite.

We use mathematical induction to show that if $h\in H_m$, $h$ is an $(\ve/3^{n-m+1})$-Nash equilibrium, and
\[\max_{i\in N}|U_i(h)-U_i(f)|<\frac{\ve}{3^{n-m+1}}.\]
If $m=1$, then $H_1=\{f\}$, and this result is obvious. Suppose that this result holds for some $m\in \{1,...,n\}$, and choose any $h'\in H_{m+1}$. If $h_m'=f_m$, then $h'\in H_m$, and thus, our claim automatically holds. Hence, we assume that $h_m'=f_m'$. Let $h$ be a strategy profile such that $h_j=h_j'$ for $j\neq m$ and $h_m=f_m$. Then, $h\in H_m$, and thus, for any $j\neq m$, $u^{mjh}$ is defined and, for a mixed strategy $g_j$ of player $j$,
\[U^{mjh}(f_m,g_j)=U(g_j,h_{-j}),\]
\[U^{mjh}(f_m',g_j)=U(g_j,h_{-j}').\]
Choose any mixed strategy $g_i'$ of player $i$ and set $g_j'=h_j'$ for all $j\neq i$. If $i=m$, then $g_j'=h_j$ for all $j\neq m$, and by the induction hypothesis,
\[U_i(g')-U_i(h)\le \frac{\ve}{3^{n-m+1}}.\]
Therefore, for some $j\neq m$,
\begin{align*}
U_i(g')-U_i(h')=&~U_i(g')-U_i(h)+U_i(h)-U_i(h')\\
\le&~U_i(g')-U_i(h)+|U^{mjh}_i(f_m,h_j)-U^{mjh}_i(f_m',h_j)|\\
<&~\frac{\ve}{3^{n-m+1}}+\frac{\ve}{3^{n-m+1}}\\
<&~\frac{\ve}{3^{n-(m+1)+1}}.
\end{align*}
Hence, we assume that $i\neq m$. Let $g_j=g_j'$ if $j\neq m$ and $g_m=h_m$. Because of the induction hypothesis,
\[U_i(g)-U_i(h)\le \frac{\ve}{3^{n-m+1}}.\]
Meanwhile,
\[|U_i(g')-U_i(g)|=|U^{mih}_i(f_m',g_i)-U^{mih}_i(f_m,g_i)|<\frac{\ve}{3^{n-m+1}},\]
\[|U_i(h')-U_i(h)|=|U^{mih}_i(f_m',h_i)-U^{mih}_i(f_m,h_i)|<\frac{\ve}{3^{n-m+1}}.\]
Therefore,
\begin{align*}
U_i(g')-U_i(h')=&~U_i(g')-U_i(g)+U_i(g)-U_i(h)+U_i(h)-U_i(h')\\
\le&~|U_i(g')-U_i(g)|+U_i(g)-U_i(h)+|U_i(h)-U_i(h')|\\
<&~\frac{\ve}{3^{n-m+1}}+\frac{\ve}{3^{n-m+1}}+\frac{\ve}{3^{n-m+1}}\\
=&~\frac{\ve}{3^{n-(m+1)+1}}.
\end{align*}
Thus, $h'$ is an $(\ve/3^{n-(m+1)+2})$-Nash equilibrium. Moreover,
\[|U_i(h')-U_i(f)|\le |U_i(h')-U_i(h)|+|U_i(h)-U_i(f)|<\frac{\ve}{3^{n-(m+1)+1}}.\]
This completes the proof of this induction. In particular, to set $m=n+1$, we have that $f'$ is an $\ve$-purification of the Nash equilibrium $f$. This completes the proof. $\blacksquare$

\section*{Declarations}
We declare that there is no conflicts of interest associated with this manuscript.

\section*{References}

\begin{description}
\item{[1]} Aumann, R. J., Katznelson, Y., Radner, R., Rosenthal, R. W., Weiss, B. Approximate Purification of Mixed Strategies. Math. Oper. Res. 8, 327-341 (1983).

\item{[2]} Balder, E. J. Generalized Equilibrium Results for Games with Incomplete Information. Math. of Oper. Res. 13, 265-276 (1988).

\item{[3]} Billingsley, P. Convergence of Probability Measures. Wiley (1999).

\item{[4]} Cotter, K. D. Correlated Equilibrium in Games with Type-Dependent Strategies. J. Econ. Theory 54, 48-68 (1991).

\item{[5]} Dudley, R. M. Real Analysis and Probability, 2nd ed. Cambridge University Press, New York (2002).

\item{[6]} Keisler, H. J., Sun, Y. Why Saturated Probability Spaces Are Necessary. Adv. Math. 221, 1584-1607 (2009).

\item{[7]} Khan, M. A., Rath, K. P. On Games with Incomplete Information and the Dvoretsky-Wald-Wolfowitz Theorem with Countable Partitions. J. Math. Econ. 45, 830-837 (2009).

\item{[8]} Khan, M. A., Rath, K. P., Sun, Y. On A Private Information Game without Pure Strategy Equilibria. J. Math. Econ. 31, 341-359 (1999).

\item{[9]} Khan, M. A., Rath, K. P., Sun, Y. The Dvoretzky-Wald-Wolfowitz Theorem and Purification in Atomless Finite-Action Games. Int. J. Game Theory 34, 91-104 (2006).

\item{[10]} Khan, M. A., Sagara, N. Relaxed Large Economies with Infinite Dimensional Commodity Spaces: the Existence of Walrasian Equilibria. J. Math. Econ. 67, 95-107 (2016).

\item{[11]} Khan, M. A., Zhang, Y. On the Existence of Pure-Strategy Equilibria in Games with Private Information: A Complete Characterization. J. Math. Econ. 50, 197-202 (2014).

\item{[12]} Milgrom, P. R., Weber, R. J. Distributional Strategies for Games with Incomplete Information. Math. Oper. Res. 10, 619-632 (1985).

\item{[13]} Nash, J. F. Equilibrium Points in N-person Games. Proc. Nat. Acad. Sci. USA 36, 48-49 (1950).

\item{[14]} von Neumann, J., Morgenstern, O. Theory of Games and Economic Behavior. Princeton University Press, Princeton (1944).

\item{[15]} Parthasarathy, K. R. Probability an Mathematical Statistics. Academic Press, New York (2014).

\item{[16]} Radner, R., Rosenthal, R. W. Private Information and Pure-Strategy Equilibria. Math. Oper. Res. 7, 401-409 (1982).

\item{[17]} Rudin, W. Real and Complex Analysis, 3rd ed. McGraw-Hill, Singapore (1987).

\item{[18]} Srivastava, S. M. A Course on Borel Sets. Springer-Verlag, New York (1988).

\item{[19]} Stinchcombe, M. B. Correlated Equilibrium Existence for Infinite Games with Type-Dependent Strategies. J. Econ. Theory 146, 638-655 (2011).

\item{[20]} Yu, C., Hosoya, Y., Maruyama, T. On the Purification of Mixed Strategies. Econ. Bull. 38, 1655-1675 (2018).

\end{description}

\end{document}